# Web 3.0 and a Decentralized Approach to Education

Sarah A. Flanery
*Department of Electrical and Computer Engineering*
*Texas A&M University*
College Station, TX, USA
sflanery@tamu.edu

Kamalesh Mohanasundar
*Department of Computer Science and Engineering*
*Texas A&M University*
College Station, TX, USA
kammohan04@tamu.edu

Christiana Chamon
*Department of Computer Science and Engineering*
*Texas A&M University*
College Station, TX, USA
cschamon@tamu.edu

Srujan D. Kotikela
*Department of Computer Science and Information Systems*
*Texas A&M University–Commerce*
Commerce, TX, USA
srujan.kotikela@tamuc.edu

Francis K. Quek
*Department of Teaching, Learning, and Culture*
*Texas A&M University*
College Station, TX, USA
quek@tamu.edu

*Abstract*—With the natural evolution of the web, the need for decentralization has rendered the current centralized education system out of date. The student does not "own" their credentials, as the only way their accomplishments are directly linked to their person and considered valuable is by verification through a stamp of an expensive, prestigious institution. However, going to a university is no longer the only way to acquire an education; open-source learning material is widely available and accessible through the internet. However, our society does not deem these methods of education as verifiable if they do not include a degree or certificate. Additionally, a valid certificate for the vast majority of open-source courses costs a few hundred dollars to obtain. The centralized nature of education inadvertently places students in underprivileged communities at a disadvantage in comparison to students in economically advantaged communities, thus a decentralized approach to education would eliminate the vast majority of such discrepancies. In the present paper, we integrate Decentralized Identity (DID) with Web 3.0 to upload credentials linked directly to the user. Each credential is appended to an Ethereum blockchain that, by design, cannot be altered once uploaded. We include DID document based access controls to display the candidate's upload and verification history. Finally, we utilize TLS protocols to provide a secure connection to the internet for ensuring non-fungibility of credentials and authentication of users.

## I. INTRODUCTION[1]

### A. Link to Web 3.0

Web 3.0 currently exists with the purpose of bringing back ownership to its users [1], [2]. To accomplish this, it takes a decentralized identity approach using blockchain technology [3]. Similar to a linked list, blockchains consist of blocks appended to a preexisting list. Each block contains information about the previous and preceding block. However, no previous blocks can be deleted or altered [4].

Information security in the present day is highly centralized; all data is controlled by singular entities such as the protocols within SSL and TLS, and users have to trust third parties to verify that their data is encrypted [5], [6]. For example, social media companies such as Twitter, Facebook, and Instagram are all operated by a centralized power: the admin assigned by each company. This system of storing information is based on Web 2.0. The basis of Web 2.0 is that a central power ensures the safety and accuracy of information online and decides what is distributed to its users. These administrators control what information is distributed to its users and which must be hidden or deleted from the platform. For these reasons, this has the potential to be a gateway to corruption, as the basis of the population's access to information are influenced by personal biases [7].

Web 3.0, due to the nature of its data structure, eliminates this problem entirely by removing the centralized power and using other methods of verification to determine the accuracy of information on the internet. The obstacle encountered when trying to incorporate Web 3.0 into education is the centralized nature of learning. In modern society, the validity of an academic credential is established by the institution itself rather than defining individual skills learned. In Web 2.0, the user does not have any ownership of the information that they post online. However, Web 3.0 resolves this issue. In the present paper, we demonstrate that the decentralized elements of Web 3.0 can be utilized in order to ensure that the accomplishments in education remain assigned to the user.

---
[1]Copied and adapted from [1]

## B. Decentralized Identity

Decentralized Identity is a form of authentication standardized by Web 3.0 that does not require a major entity to validate the user [?]. In the current state of the internet, the user's identity is tied to the centralized source. For example, most academic applications involve having to utilize an educational account that is created by an external provider. If these external providers were to become deactivated, these personal accounts would be displaced. Decentralized identity attempts to resolve this by using an Ethereum blockchain to append blocks tied to each user's specific decentralized identifier (DID). Centralized figures are unable to regulate DID users, and this provides ownership of the user's access and possession of online data. Increased ownership of information, similar to the basis of Web 3.0, has rendered DID a normalized method of self-identity. For example, the European Union (EU) has been trying to adopt DID in order to replace other forms of government ID such as passports and driver's licenses [8]. Citizens can not only have all their necessary identification in a singular place, but also as a method to decrease potential government corruption.

DID is generated using asymmetric cryptography, consisting of a private key pair and a hash that is automatically generated in Ethereum. The DID document containing the identity of the user is then verified by either an outside entity or the user's self. In the latter case, biometric login such as fingerprinting and face ID is often used, as it cannot be easily replicated by outside persons [9]. Once verified, the DID document, such as the one displayed in Fig. 1, is appended to a distributed ledger, often a blockchain. However, blockchain architecture is not required to perform this. Decentralized Identifiers allow users to take back ownership of their online credentials, as there is no centralized figure managing the ecosystem.

## C. Ownership of Learning

*1) Validity:* In the current economic system, people are rewarded currency in exchange for being productive members of society, thus there is a will for contribution. Such contributions are derived from knowledge and skills, which are paved in multiple pathways such as college courses, makerspace classes, and online learning. People summarize into one page the knowledge and skills they've acquired as a result of their learned experiences, and they create profiles on LinkedIn to keep a record of their employment history, certifications, and skills. Once knowledge and skills are acquired, they are permanent and cannot be transferred, duplicated, or stolen; users have complete ownership of their knowledge and skills, as they are non-fungible.

The web's current influence on self-promotion based on knowledge and skills allows for self-proclamation of such skills, or self-credentialing. On LinkedIn, anyone can add a skill to their profile, and their connections can endorse without verification of the knowledge of the skill. One way job interviewers verify that applicants are proficient in the skills that they list involves asking technical questions to ensure that applicants have enough general knowledge to be able to perform the required tasks. Additionally, the applicant may not have known that they were "lying" on their resume, as they might have surface-level knowledge of such a skill, but another applicant may have granular knowledge of that same exact skill. Each institution has its own criteria for what constitutes successful completion of a skill, often in the form of a letter grade reflecting the range of percentage points a student acquired over the semester. However, the letter grade alone does not determine the extent of the material learned. One instructor may have been lazy with grading the students homework, allowing it to be considered a "blow off class" or "easy A," while another instructor created a rigorous rubric and curriculum where a student earns a C. The end result is that these two students received different grades that did not linearly represent their knowledge and skills, as the person with the C has acquired more knowledge than the student who received an A with little to no effort.

The problem with this lack of standardization and uniformity poses the following questions: what exactly do these credentials capture? What makes a resume credential verified? How is the validity of a skill properly verified given not all instructors and institutions provide the same depth of experience? Can it be determined that the credential was not forced or given undeservedly? How do we account for these short-term factors that greatly impact the modern workforce in the long term? Our pilot study seeks to propose a solution to this validity problem.

*2) Trust:*

*a) Institutionalization:* People pay high tuition prices to take courses from a university that is supposed to provide knowledge and skills, and they are awarded a piece of paper (diploma) which states that they have satisfied a certain curriculum of courses. Students who don't finish college don't have a piece of paper to prove their knowledge, despite their knowledge not being lost. Rather than a coarse credential representing an aggregate set of skills, these skills can instead be considered as micro credentials. An entity that attempts to attack this issue is BCDiploma [10]. This service is used by companies and universities to digitize their diplomas, credentials, and certificates in a secure manner that cannot be fabricated due to the blockchain data structure. However, a problem with institutionalized learning is that the institution is the single point of failure. If the data in the institution gets attacked, then everyone (students, staff, faculty, alumni) who is affiliated with the institution as well as the reputation of the institution's name would be invalidated. This raises the question: what qualities does a university such as Harvard possess that renders it prestigious?

Prestige is what makes the general population have so much faith in the quality of Harvard education. This is established by reputation, scarcity, and monetary value. These elements are linked: Harvard has its reputation due to its selectivity in admissions. Similar to economics, there is a linear correlation between the monetary value of a resource with respect to its scarcity. Despite the cost of education and its intrinsic value not having a linear correlation, we as a society view it as

```
DID                                                                                              Alias
did:we ┌─did:ethr:0x025f2c19148f0afb66ef9f3c7cc65464ba2d4e9339784d1a6a7239c059e1f75014─
did:et │ {
did:we │   "did": "did:ethr:0x025f2c19148f0afb66ef9f3c7cc65464ba2d4e9339784d1a6a7239c059e1f75014",
did:et │   "provider": "did:ethr",
did:et │   "alias": "Kamalesh Mohanasundar",
did:et │   "controllerKeyId": "045f2c19148f0afb66ef9f3c7cc65464ba2d4e9339784d1a6a7239c059e1f750149c8a5997fe2f713eb8d31db93b5b5e8716332
did:et │ fb17a9333643d3fd3f5748cbfc0",
did:et │   "keys": [
did:et │     {
did:et │       "kid": "045f2c19148f0afb66ef9f3c7cc65464ba2d4e9339784d1a6a7239c059e1f750149c8a5997fe2f713eb8d31db93b5b5e8716332fb17a933
did:et │ 3643d3fd3f5748cbfc0",
did:et │       "kms": "local",
did:et │       "type": "Secp256k1",
did:et │       "publicKeyHex": "045f2c19148f0afb66ef9f3c7cc65464ba2d4e9339784d1a6a7239c059e1f750149c8a5997fe2f713eb8d31db93b5b5e871633
did:et │ 2fb17a9333643d3fd3f5748cbfc0",
               "meta": {
                 "algorithms": [
                   "ES256K",
                   "ES256K-R",
                   "eth_signTransaction",
```

Fig. 1. Typical DID document for alias "Kamalesh Mohanasundar" consisting of the domain of the ledger ("did:ethr"), the user's specifically-assigned DID (given by "did"), and respective public key (given by "publicKeyHex").

such, especially when having to browse through thousands of resumes; applicants with prestigious institutions on their resume have a major advantage. This same logic applies to independent online certification. Similar to the approach taken with MOOCs (see Section I-C2b), a lot of the value within the education system has a direct correlation with the cost required to attend, rendering it very difficult for low income students to attend these schools [12]. In order to increase opportunities for students of all backgrounds, we raise the following question: how can we verify the prestige of an institution without monetary means?

With the internet becoming more accessible to the general population, it allows for free education resources to be a regular phenomenon. People are free to watch YouTube tutorials and other forms of free education at any point in the day. This, however, comes with its own set of challenges in verifying the credentials earned through these services.

*b) Massive Online Open Courses:* Massive Open Online Courses (MOOCs) are available for public use, and they are tied to multiple credible institutions such as Harvard and Google. EdX, for example, is a third party source that requires a monthly/per course fee of $̃250. Once the transaction is made, you will be allowed to earn a certificate for the course. The majority of the courses also are available for free; however, they do not come with a certificate, thus making it difficult for employers to verify whether or not the client mastered the skills gathered from the course. Aside from these institutions having pre-established credibility, the value of the skills is determined not by the skills acquired by the user, but rather if they are willing to pay the fee. This same logic applies to our modern education system, where money is a major factor in what makes these institutions "valuable" and scarce. Our pilot study seeks to nullify this issue by eliminating the need for a paid certificate, and instead having an issuer upload credentials.

The rest of this paper is organized as follows. Section II describes the methodology used in our pilot study, Section III provides a demonstration thereof, and Section IV concludes this paper.

## II. METHODOLOGY

For this pilot study, we create multiple users, generate DIDs for them, and append them to an Ethereum blockchain. We then create a verifier to assign multiple credentials that are tied to the student (user). For each user, we group credentials that are similar to one another (e.g. C++ Course, Python Project, Software Intern) through the presentation feature, which allows users to create a digital portfolio of their related works. Finally, we create DIDs for employers and institutions to be able to pursue the users as candidates with presentations that match the required skills for the desired job position or promotion.

## III. DEMONSTRATION

### A. Setup

We used Veramo to issue DIDs and credentials because of its open-source framework that facilitates creating and managing digital identities [13], [14]. By taking advantage of Veramo's framework and modular design, we are able to incorporate the decentralized identity management and credential verification functionalities to successfully generate DIDs (see Section I-B).

We also linked Veramo to OpenSSL for the purpose of providing an extra layer of security in comparison to using Veramo alone. Using the OpenSSL GitHub repository, we were able to establish secure connections between web servers [11]. External security in a decentralized environment ensures that operations between systems are not maliciously infiltrated.

Integration into a demo web app involved installing Node dependencies such as "yarn" and creating the agent within a nodeJS app. Because the plugins Veramo uses are all public and standardized, any verifier can resolve DIDs.

*B. Integration*

To simulate a real use case, we integrated Veramo into a hypothetical university by creating sample users and generating a respective DID to store in an Ethereum blockchain. To demonstrate the application of decentralized identifiers using the Veramo framework, we initiated the process by entering the command "veramo", which showed a list of features and commands. To simulate a single user, we created a sample DID by entering the command "veramo did create", and we were then given the option to choose between five different types of DID generators: Ethereum, Ethereum Goerli, Web, Key-based, and PKH. We chose Ethereum specifically for the purpose of allowing all users to be appended to its blockchain, as described in Sections I-B and II. Adding the DID to the Ethereum blockchain rendered a display as shown in Fig. 2.

To append credentials to a DID, we entered "veramo credential create" and selected Ethereum. To simulate a credentialer, we input the user's assigned DID, followed by the issuer's DID and the credential being issued. The respective user interface is shown in Fig. 3.

In order to simulate the search of a specific user or credentials from an employer's or institution's point of view, we utilized the command "explore". Fig. 4 displays the menu for the aforementioned options. "Managed identifiers" provided a list of users and their associated DIDs. When a user was selected, their entire DID document was displayed, i.e. the issuer, subject, and credential details as displayed in Fig. 5. The "Messages" feature allowed users to send messages to one another using their specific DID. The "Credentials" option had a similar function to "Managed identifiers"; however, only the credentials were displayed, as shown in Fig. 6. Finally, the "Presentation" option enabled the employer to see a user's chosen compilation of related skills, as seen in Fig. 7. When selected, Veramo displayed a list of user-defined presentations containing a compilation of professional and personal projects. We as the user selected credentials that we believed to be connected to one another, and then we placed them in a singular repository. This allowed the employers and institutions to see multiple credentials at once, rather than searching for them individually.

## IV. CONCLUSION

The centralized nature of education naturally puts underrepresented groups at an economic disadvantage due to the linear correlation between tuition costs and the prestigious value of an institution. We demonstrated in our pilot study that decentralized education can be accomplished by using decentralized identifiers and the foundations of Web 3.0. The Veramo demonstration displayed people of all educational backgrounds and their unique skill sets, which were verified by other users. Credentials were then compiled into digital portfolios for potential employers, creating an ecosystem in which monetary certificates are not required to show proof of knowledge.

This pilot can continue its development with the addition of an authentication scheme involving biometrics for the purpose of tying the users' DID to an element of their person that cannot be replicated. Another improvement would involve the addition of a data aggregation scheme such that employers and institutions can have their search results narrowed down by relevance to the job position, how frequently the users upload their credentials, and the newest credential entries, for once a large number of users join the ecosystem, searching for specific projects will no longer be sufficient.

```
PS C:\Users\sarsa\Documents\GitHub\veramo-nodejs-tutorial> veramo did create
(node:15212) ExperimentalWarning: Importing JSON modules is an experimental feature and might change at any time
(Use `node --trace-warnings ...` to show where the warning was created)
? Select identifier provider did:ethr
? Select key management system local
? Enter alias Luke Skywalker
```

| provider | alias | did |
|---|---|---|
| did:ethr | Luke Skywalker | did:ethr:0x02c81c2097947f5e072c6121325f1cb02221df2a14eb7e1ec1c4b28722ae2951bf |

```
PS C:\Users\sarsa\Documents\GitHub\veramo-nodejs-tutorial>
```

Fig. 2. Example of a generated User Decentralized Identifier (shown as "did") using Ethereum ("ethr"). The user chooses an alias to display their preferred name. The respective DID is created using an Ethereum-generated hash.

```
● PS C:\Users\sarsa\Documents\GitHub\veramo-nodejs-tutorial> veramo credential
  (node:19700) ExperimentalWarning: Importing JSON modules is an experimental feature and might change at any time
  (Use `node --trace-warnings ...` to show where the warning was created)
  ? Credential proofFormat EthereumEip712Signature2021
  ? Issuer DID did:ethr:0x02072c398d8e1320f3e0cab7b5d2ecf1bf38805dede917a50926ff238ae1ea1737 Sarah Flanery
  ? Subject DID did:ethr:0x02c81c2097947f5e072c6121325f1cb02221df2a14eb7e1ec1c4b28722ae2951bf
  ? Credential Type (Python Course, Profile)
  ? Claim Type Mastery of Python Programming Language
  ? Claim Value A+
  ? Is the credential revocable? No
```

Fig. 3. The credentialer inputs their DID (Issuer DID), their student's DID (subject DID), and lists details of the assigned credential which includes the Credential Type, its "value" to demonstrate the user performance, and whether or not the credential can be revoked.

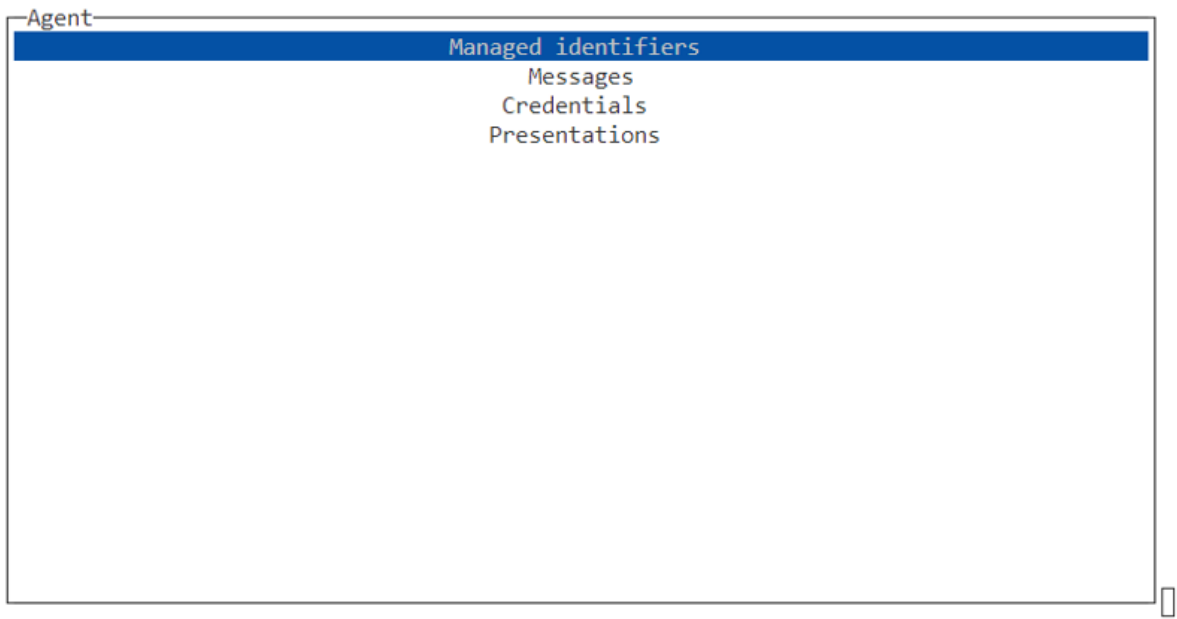

Fig. 4. Veramo Explore menu containing "Managed identifiers", "Messages", "Credentials", and "Presentations".

Fig. 5. "Managed identifiers" feature from the Veramo Explore menu (see Fig. 4). Decentralized identifiers are noted by their provider ("web" or "ethr") and the associated hash. Preferred names are listed to the right of each respective DID, under "Alias".

Fig. 6. "Credentials" feature from the Veramo Explore menu (see Fig. 4). All credentials are listed with their creation, associated "Type", the issuer DID (shown under "From"), and the user who owns the credentials (shown under "To").

Fig. 7. Example "Presentation" from the Veramo Explore menu (see Fig. 4), containing two similar credentials. When employers view the presentation, they can all see relevant credentials compiled to a single space. The presentation contains the Ethereum "hash", its assigned name "tag", proof of each highlighted credential, and their respective verification information.